\documentclass[aps,prl,preprint,superscriptaddress,showpacs,nofootinbib,preprintnumbers]{revtex4-1}
\usepackage{amsmath}
\usepackage{graphicx}
\usepackage{amssymb}
\usepackage{amsthm}
\usepackage{dsfont}
\usepackage{verbatim}

\DeclareMathOperator{\Tr}{\rm{Tr}}

\begin{document}
\title{Entanglement and Particle Identity: A Unifying Approach}
\author{A.P. Balachandran}
\email[]{bal@phy.syr.edu}
\affiliation{Institute of Mathematical Sciences, CIT Campus, Taramani,
Chennai
600113, India}
\affiliation{Physics Department, Syracuse University, Syracuse, NY,
13244-1130,
USA}

\author{T.R. Govindarajan}
\email[]{trg@imsc.res.in}
\affiliation{Institute of Mathematical Sciences, CIT Campus, Taramani,
Chennai
600113, India}

\affiliation{Chennai Mathematical Institute, H1, SIPCOT IT Park,
Kelambakkam,
Siruseri 603103, India}

\author{Amilcar R. de Queiroz}
\email[]{amilcarq@unb.br}
\affiliation{Instituto de Fisica, Universidade de Brasilia, Caixa Postal
04455,
70919-970, Brasilia, DF, Brazil}
\altaffiliation{Institute of Mathematical Sciences, CIT Campus, Taramani,
Chennai 600113, India}

\author{A.F. Reyes-Lega}
\email[]{anreyes@uniandes.edu.co}
\affiliation{Departamento de F\'isica, Universidad de los Andes, A.A.
4976,
Bogot\'a D.C., Colombia}
\altaffiliation{Institute of Mathematical Sciences, CIT Campus, Taramani,
Chennai 600113, India}

\date{\today}

\begin{abstract}
It has been known for some years that entanglement entropy obtained from partial trace does not provide the correct
entanglement measure when applied to systems of identical particles. Several criteria have been proposed that have the
drawback of being different according to whether one is dealing with fermions, bosons or distinguishable particles.
In this Letter, we give a precise and mathematically natural answer to this
problem. Our approach is based on the use of the more general idea of
restriction of states to subalgebras. It leads to a novel approach to entanglement, suitable to be used in general
quantum systems and specially in systems of identical particles. This settles some recent controversy regarding
entanglement for identical particles. The prospects for applications of our criteria are wide-ranging, from spin chains
in condensed matter to entropy  of black holes.
\end{abstract}
 \maketitle
\section{Introduction}
\label{sec:1} The study of subsystems of a quantum system is of paramount importance in many branches of physics. In
quantum information it enters in the analysis of local operations performed by different parties of a multipartite
system. In statistical physics it enters in the very definition of the different ensembles since this involves
considering a given physical system as embedded in a bigger one.  In the physics of black holes, the distinction
between accessible and inaccessible regions of space-time plays a crucial role for the study of black hole entropy.
Indeed as pointed out in \cite{Sorkin1998}, the coupling from outside to inside the horizon is very strong, while the
reverse coupling is nonexistent!   In all these situations partial trace is the preferred tool to extract physical
properties of the given subsystems. Nevertheless, it is well-known that in some cases of great physical interest like
systems of identical particles the use of partial trace leads to contradictory results.

In this Letter we provide a resolution of such contradictions which turns out to be of general application. We show
that, by treating observables and states on an equal footing, a generalized notion of entanglement emerges. A relevant
consequence is that the entanglement measure that naturally arises in this algebraic approach is shown to be easily
computed. Our approach thus opens up a wide range of applications, from condensed matter systems, like spin chains and
anyonic models, to  black hole physics.

For bipartite systems contradictory results due to partial trace are explicitly seen to appear in the computation of
entanglement measure for identical particles systems. In spite of the numerous  efforts  to achieve a satisfactory
understanding of entanglement for systems of identical particles, there is no  general agreement on the appropriate
generalization of concepts valid for non-identical
constituents~\cite{Adhikari2009,Amico2008,Eckert2002,Ghirardi2004,Schliemann2001,Tichy2011,grabowski2011entanglement}.
That is because many concepts  are usually only discussed in the context of quantum
 systems  for which the Hilbert space $\mathcal H$ is a simple tensor
product with no additional structure like, for example,  $\mathcal H=\mathcal H_A \otimes\mathcal H_B$. In this case,
the partial trace $\rho_A=\Tr_B |\psi\rangle\langle \psi|$ for $|\psi\rangle \in \mathcal H$ to obtain the reduced
density matrix has a good physical meaning: it corresponds to observations only on the subsystem $A$.

In contrast, the Hilbert space of a system of $N$ identical bosons (fermions) is given by the symmetric (antisymmetric)
$N$-fold tensor product of the single-particle spaces. The consequence is that any multi-particle state contains
\emph{intrinsic} correlations between subsystems due to quantum indistinguishability. This, in turn, forces a departure
from the straightforward application of  entanglement-related concepts like singular value decomposition (SVD), Schmidt
rank or  entanglement entropy.

We propose here an approach to the study of entanglement where the notion of \emph{partial trace} is replaced by the
more appropriate notion of \emph{restriction of a state to a
subalgebra}\cite{Barnum2004,*fritz2010operator,*derkacz2012}. This approach  is based on the well
established GNS construction\cite{haag1996local}. It allows us to meaningfully treat entanglement of identical and
non-identical particles on an equal footing, without the need to resort to different criteria according to the case
under study.

 The usefulness of our approach will be displayed in three explicit simple
examples (for more examples see \cite{Balachandran2012a}). In particular we obtain a vanishing  von Neumann entropy of
a fermionic or a bosonic state containing the least possible amount of correlations. We believe that this settles an
issue that has caused a lot of confusion regarding the use of von Neumann entropy as a measure of entanglement for
identical particles \cite{Tichy2011,Paskauskas2001, Wiseman2003}.

\section{The GNS Construction}
A general quantum system is usually described in terms of a Hilbert space $\mathcal H$ and linear operators acting
thereon. Physical \emph{observables} correspond to  self-adjoint operators ($\mathcal O\equiv\mathcal
O^\dagger:\mathcal H\rightarrow\mathcal H$). The probabilistic character of the theory is based on the notion of
\emph{state}, from which probabilities and expectation values can be computed. Generically,  a state is described in
terms of  a \emph{density matrix} $\rho:\mathcal H \rightarrow \mathcal H$, a linear map satisfying  $\Tr \rho =1$
(normalization), $\rho^\dagger=\rho$ (self-adjointness) and $\rho \geq 0$ (positivity). For \emph{pure} states, the
additional condition $\rho^2=\rho$ is required, so that $\rho$ is of the form $|\psi\rangle\langle\psi|$ for some
normalized vector $|\psi\rangle\in\mathcal H$.

Since the expectation value of an observable $\mathcal O$ is defined by $\langle \mathcal O\rangle_\rho= \Tr (\rho
\,\mathcal O)$, we can equivalently regard $\rho$ as a  \emph{linear functional} $\mathcal O\mapsto \langle\mathcal
O\rangle_\rho$ from the space of  operators to $\mathds C$. Moreover, since the space of all (bounded) operators on
$\mathcal H$ forms an algebra $\mathcal L(\mathcal H)$, it is possible to give a formulation of quantum physics which
does not \emph{a priori} make use of Hilbert spaces. Such a formulation was initially envisaged by von Neumann. The
formulation due to Gel'fand and Naimark and further developed by Segal (GNS construction) led to the notion of an
``abstract algebra of physical observables", or $C^*$-algebra. This construction (explained below) has played a very
important role in quantum field theory \cite{haag1996local} and statistical mechanics \cite{bratteli}. We propose to
show that this approach is also very well-suited to deal with the problem described in the introduction.

We thus consider an abstract algebra $\mathcal A$ (playing the role of $\mathcal L(\mathcal H)$ above) that represents
the physical observables. Since these observables are (not yet) acting on any Hilbert space, an abstract notion for the
adjoint of an operator is required. This is provided by an operation (``involution'') $\alpha\mapsto \alpha^*$. The
algebra is assumed to contain an identity $\mathds 1_{\mathcal A}$ and to be closed under products, linear combinations
and under the involution. In this context, a \emph{state} is defined as a linear functional $\omega: \mathcal A
\rightarrow \mathds C$. Again, since there is no Hilbert space, no density matrix appears at this stage. But from the
interpretation of $\omega(\alpha)\equiv \langle \alpha\rangle_\omega$ as the expectation value of $\alpha$, the
conditions of normalization $\omega(\mathds 1_\mathcal A)=1$, reality $\omega(\alpha^*)=\overline{\omega(\alpha)}$ and
positivity $\omega (\alpha^* \alpha)\geq 0$ (for any $\alpha\in \mathcal A$) are physically motivated properties that
any state $\omega$ must, by definition, satisfy.

Given a quantum system defined by an algebra $\mathcal A$ and a state $\omega$, how do we recover the usual Hilbert
space on which the algebra elements act as linear operators? Since $\mathcal A$ is an algebra, it is in particular a
vector space, denoted here as $\hat{\mathcal{A}}$. Elements $\alpha \in \mathcal A$ regarded as elements of the vector
space $\hat{\mathcal A}$ are written as $|\alpha \rangle$. Then, $\beta\in\mathcal A$ will act on
$|\alpha\rangle\in\hat{\mathcal A}$ as a linear operator by $\beta |\alpha\rangle :=|\beta\alpha\rangle$. A similar
construction occurs when we study the \emph{regular representation} of a group through its action on its group
algebra~\cite{Balachandran2010}.

In order for the vector space $\hat{\mathcal A}$ to become a Hilbert space, an inner product is required. If we set
$\langle \alpha |\beta \rangle = \omega (\alpha^* \beta)$, we obtain almost all properties of an inner product. In
fact, reality and positivity can be used to show that $\langle \beta|\alpha\rangle=\overline{\langle \alpha |\beta
\rangle}$ and also that $\langle \alpha| \alpha \rangle \geq 0$. But it can happen that $\langle \alpha| \alpha \rangle
= 0$ for some non-zero elements $\alpha$. That is,  there could be a null space $\widehat{\mathcal{N}}_\omega$ of zero
norm vectors: $\widehat{\mathcal{N}}_\omega=\lbrace |\alpha \rangle \in \hat{\mathcal A} \,|\,\omega (\alpha^*
\alpha)=0\rbrace$. The solution to this problem is obtained by considering the \emph{quotient vector space}
$\hat{\mathcal A}/ \widehat{\mathcal{N}}_\omega$. Its elements are \emph{equivalence classes} $|[\alpha]\rangle$,  with
$|[\alpha]\rangle$ equivalent to $|[\beta]\rangle$ precisely when $\alpha -\beta \in \widehat{\mathcal{N}}_\omega$. In
particular, if $\alpha\in \widehat{\mathcal{N}}_\omega$, then $|[\alpha]\rangle =0$. The space $\hat{\mathcal A}/
\widehat{\mathcal{N}}_\omega$ has now a well-defined scalar product given by
\begin{equation}
\label{eq:inner}
\langle[\alpha] |[\beta]\rangle = \omega (\alpha^* \beta ),
\end{equation}
independently of the choice of $\alpha$ from $[\alpha]$ and with no non-zero
vectors of zero norm. Its closure is the  GNS Hilbert space $\mathcal
H_\omega$. In this way, one obtains  a representation $\pi_\omega$ of $\mathcal A$ on $\mathcal H_\omega$  by linear
 operators~\cite{haag1996local,Landi2008}: $\pi_\omega(\alpha)|[\beta]\rangle=|[\alpha\beta]\rangle$.

\textbf{Partial Trace as Restriction - } Consider  a  bipartite system $\mathcal H=\mathcal H_A\otimes \mathcal H_B$,
with a density matrix $\rho$. The description of $\mathcal H_A$ as a \emph{subsystem} involves the reduced density
matrix  $\rho_A$, obtained through partial tracing over $B$. Using the language of algebras and states, we observe that
the algebra corresponding to the joint system $AB$ is given by $\mathcal A= \mathcal L(\mathcal H)$. Expectation values
are computed using the state $\omega_\rho$ induced by $\rho$, $\langle\mathcal O\rangle_\rho\equiv \omega_\rho(\mathcal
O)\equiv \Tr_{\mathcal H}(\rho\, \mathcal O)$. Corresponding to subsystem $A$, we can consider the subalgebra $\mathcal
A_0$ of ``local'' operators of the form $K\otimes\mathds{1}_B$, for $K$ an observable on $\mathcal H_A$.  We can then
define a state $\omega_{\rho,0}: \mathcal A_0\rightarrow \mathds C$ which is the \emph{restriction}
$\omega_\rho|_{\mathcal{A}_0}$ of $\omega_\rho$ to $\mathcal A_0$ defined by
$\omega_{\rho,0}(\alpha)=\omega_\rho(\alpha)$ if $\alpha\in\mathcal{A}_0$.

Now we observe that the reduced density matrix $\rho_A$, obtained by partial tracing, gives rise to a state on
subsystem $A$ that is precisely the restriction of $\omega_\rho$ to $\mathcal{A}_0$:
\begin{equation}
\label{eq:restriction}
\omega_{\rho,0}(K\otimes \mathds 1_B)\equiv\Tr_{\mathcal H_{\mathcal A}}(\rho_A\,K).
\end{equation}
Hence,  partial trace and restriction give the same answer in this case. The importance of this observation lies in the
fact that when $\mathcal H$ is not of the form of a `simple tensor product', partial trace is not a suitable operation.
In contrast, if the system is described in terms of a state $\omega_\rho$ on an algebra $\mathcal A$, it is still
sensible to describe a subsystem in terms of a corresponding subalgebra ${\mathcal A}_0$ and  of the restriction
$\omega_{\rho,0}$ of $\omega_\rho$ to ${\mathcal A}_0$. The GNS theory is well-suited for the study of
$\omega_{\rho,0}$ for general algebras ${\mathcal A}_0\subseteq \mathcal A$.

\textbf{von Neumann Entropy -} The representation $\pi_\omega$ is in general reducible. This  means that $\mathcal
H_\omega$ can be decomposed into a direct sum of irreducible spaces: $\mathcal H_\omega= \bigoplus_i \mathcal H_i$,
where
 $\pi_\omega (\alpha) \mathcal H_i\subseteq\mathcal H_i$ for all $\alpha\in \mathcal A$.
 Let $P_i:\mathcal H_\omega \rightarrow\mathcal H_i$
be the corresponding orthogonal projectors. These projectors can be used to construct a density matrix $\rho_\omega$ on
the GNS space $\mathcal H_\omega$ that yields the same  expectation values as  the original state $\omega$. The von
Neumann entropy of $\rho_\omega$ can then be evaluated in the standard way. The construction of $\rho_\omega$ goes as
follows.

First, we observe that the identity $\mathds 1_\mathcal A$   of $\mathcal A$ satisfies $\mathds 1_\mathcal A
\alpha=\alpha$ for all $\alpha\in \mathcal A$, as well as $\mathds 1^*_{\mathcal A}=\mathds 1_\mathcal A$. This,
together with (\ref{eq:inner}), implies $\omega(\alpha)=\langle[\mathds 1_\mathcal A]|[\alpha]\rangle $. Since the
linear operator $\pi_\omega(\alpha)$ is defined by $\pi_\omega(\alpha)|[\beta]\rangle= |[\alpha \beta]\rangle$, we know
that $|[\alpha]\rangle= \pi_\omega(\alpha)|[\mathds 1_\mathcal A ]\rangle$. It follows that $\omega(\alpha)=\langle
[\mathds 1_{\mathcal A}]|\pi_\omega(\alpha) |[\mathds 1_{\mathcal A}] \rangle$. Using $|[\mathds 1_{\mathcal A}]\rangle
= \sum_i P_i |[\mathds 1_{\mathcal A}]\rangle$, $\pi_\omega(\alpha)=\sum_i P_i \pi_\omega(\alpha) P_i$ and from the
orthogonality of the projectors, one obtains $\omega(\alpha)=\Tr_{\mathcal H_\omega}\left( \rho_\omega \,
\pi_\omega(\alpha) \right)$, where $\rho_\omega= \sum_i P_i\,|[\mathds 1_{\mathcal A}]\rangle\langle [\mathds
1_{\mathcal A}] | P_i $. The von Neumann entropy of $\rho_\omega$ is then given by $S(\rho_\omega)=-\sum_i \mu_i \log_2
\mu_i$, where $\mu_i=\|P_i\,|[\mathds 1_{\mathcal A}]\rangle\|^2$.

The crucial fact is that $\omega$ is \emph{pure} if and only  if the representation $\pi_\omega$ is \emph{irreducible}.
In particular, the von Neumann entropy of $\omega$,  $S(\omega)\equiv S(\rho_\omega)$, is zero if and only if $\mathcal
H_\omega$ is irreducible. {\it{This property depends on \emph{both} the algebra $\mathcal A$ and the state $\omega$}}.

Consider now a subalgebra $\mathcal A_0\subset \mathcal A$ of  $\mathcal A$. Let $\omega_0$ denote the
\emph{restriction} to $\mathcal A_0$ of a pure state $\omega$ on  $\mathcal A$ \cite{*[{Importance of focusing on
subsystems has also been emphasized by }][{}] Barnum2004}. We can apply the GNS construction to the pair $(\mathcal
A_0, \omega_0)$ and  use the von Neumann entropy of $\omega_0$ to study the entanglement emergent from restriction.

\textbf{Bipartite Entanglement from GNS - }
 We now illustrate how to apply the GNS construction to entanglement. Consider
$\mathcal H=\mathcal H_A\otimes\mathcal H_B\equiv\mathds C^2\otimes \mathds C^2$. The algebra of the full system is
$\mathcal{A}=M_2(\mathds{C})\otimes M_2(\mathds{C})$. Let us consider the normalized state vector ($0<\lambda<1$):
$|\psi_\lambda\rangle=\sqrt{\lambda} |+,-\rangle +\sqrt{(1-\lambda)} |-,+\rangle\,$, with corresponding state $\omega$
on the algebra $\mathcal A$: $\omega(\mathcal O)=\langle\psi_\lambda|\mathcal O |\psi_\lambda\rangle$, $\mathcal O \in
\mathcal A$.

\emph{Entanglement} of $|\psi_\lambda\rangle$ is to be understood in terms of correlations between ``local''
measurements performed separately on subsystems $A$ and $B$. Measurements performed on $A$  correspond to the
restriction $\omega_0= \omega\mid_{\mathcal A_0}$ of $\omega$ to the subalgebra $\mathcal A_0\subset \mathcal A$
generated by elements of the form $\alpha\otimes \mathds 1_2$, with $\alpha\in M_2(\mathds{C})$. We obtain
$\omega_0(\alpha\otimes\mathds 1_2)=\langle \psi_\lambda| \alpha\otimes \mathds 1_2 |\psi_\lambda\rangle=
\lambda\langle +|\alpha|+\rangle +(1-\lambda)\langle -|\alpha|-\rangle$. In accordance with (\ref{eq:restriction}), we
have $\omega_0(\alpha\otimes\mathds 1_2) =\Tr_{\mathds C^2} (\rho_A \alpha)$, where $\rho_A=\Tr_B
|\psi_\lambda\rangle\langle\psi_\lambda|$, namely,
\begin{equation}
\rho_A=\left(\begin{array}{cc} \lambda & 0\\ 0 & 1-\lambda\end{array}\right).
\end{equation}
Now we perform the GNS construction based on the algebra $\mathcal A_0 \cong M_2(\mathds C)$ and  the state $\omega_0$.
These are the data needed to describe subsystem $A$. For $\alpha\in M_2(\mathds C)$, we have $\omega_0(\alpha)=\lambda
\alpha_{11} +(1-\lambda)\alpha_{22}$. Now we consider $\mathcal A_0$ as a \emph{vector space}. This is just the
assertion that $M_2(\mathds C)$ is, by itself, a vector space.  From the explicit form of $\omega_0$, one readily
concludes that, as long as $0<\lambda<1$, there are no null states. This means that the GNS space $\mathcal
H_{\omega_0}$ is just the \emph{four dimensional} space of $2\times 2$ matrices, endowed with the inner product
$\langle \alpha|\beta\rangle= \omega_0(\alpha^\dagger \beta).$ We can consider a basis of four $2\times 2$ matrices
defined as $e_{ij}= |i\rangle\langle j|$ for $i,j\in\lbrace1,2\rbrace$, where $|1\rangle\equiv |+\rangle$ and
$|2\rangle\equiv |-\rangle$. Then, for example, $\langle e_{11}|e_{11}\rangle = \lambda$ and $\langle
e_{22}|e_{22}\rangle = 1-\lambda$. With an appropriate normalization and ordering of this basis, one checks that the
operator corresponding to $\alpha\in \mathcal A_0$ is the $4\times 4$ matrix $\pi_{\omega_0}(\alpha)=
\left(\begin{array}{cc} \alpha & 0\\ 0& \alpha\end{array}\right)$, showing in an explicit way that the representation
is reducible (the GNS-space splits as $\mathcal H_{\omega_0}= \mathds C^2\oplus \mathds C^2$). Following the
prescription described above one  obtains, for the density matrix,  $\rho_{\omega_0}=
\mbox{diag}\lbrace\lambda,0,0,1-\lambda\rbrace$. The identity $\omega_0(\alpha)=\Tr_{\mathcal
H_{\omega,0}}(\rho_{\omega_0} \pi_{\omega_0}(\alpha))$ is readily checked.

From the explicit form of $\rho_{\omega_0}$ we conclude that the entropy of the \emph{restricted} state is
$S(\omega_0)= -\lambda \log_2 \lambda-(1- \lambda) \log_2 (1-\lambda)$. This is precisely the entropy  of the
\emph{reduced} density matrix $\rho_A$ obtained by partial tracing. Recalling that a (pure) state of the full system is
entangled with respect to a bipartition into subsystems if and only if $S(\rho_A)>0$, we  have thus verified that our
method reproduces the standard results in the case of bipartite systems. This is in fact a general result:

 \emph{For bipartite systems of the form} $\mathcal H_A\otimes\mathcal H_B$ \emph{(pure case), the GNS construction yields a vanishing entropy
 for the restricted state precisely when the original state
of the full-system is separable. Moreover, in the case of entangled states, the entropy computed via the GNS construction
coincides with the von Neumann entropy of the reduced density matrix computed via partial trace and can therefore be used as an entanglement measure.}

We remark that, in the pure case, entanglement can also be characterized by the impossibility of writing the state
$\omega$ as a product state. That is, if $\omega$ is of the form  $\omega(\alpha\otimes\beta)=
\omega_A(\alpha)\omega_B(\beta)$ for $\alpha$ ($\beta$) any observable on subsystem $A$ ($B$) and $\omega_A$,
$\omega_B$ states on the corresponding subsystems, then $\omega$ is a \emph{product}, or \emph{separable} state, and it
is not entangled. This observation makes clear that entanglement for \emph{mixed} states can also be studied using our
approach: If a mixed state $\omega_{\mbox{\tiny m}}$ can be written as a convex combination of product states, then it
is called separable, otherwise it is called entangled.
\section{Systems of Identical Particles}
In the case of identical particles, the Hilbert space of the system is not anymore of the tensor product form.
Therefore, the treatment of subsystems using partial trace becomes problematic. In contrast, in our approach all that
is needed to describe a subsystem is the specification of a subalgebra corresponding to the subsystem. Then, the
restriction of the original state to the subalgebra provides a physically motivated generalization of the concept of
partial trace, the latter not being sensible anymore. Applying the GNS construction to the restricted state, we can
study the entropy emerging from the restriction and use it as a generalized measure of entanglement.

Let $\mathcal H^{(1)}=\mathds C^d$ be the Hilbert space of a one-particle system. The $k$-particle Hilbert space
$\mathcal{H}^{(k)}$ for bosons (fermions) is the symmetrized (antisymmetrized) $k$-fold tensor product of
$\mathcal{H}^{(1)}$. To any  one-particle observable $A^{(1)}$ on $\mathcal{H}^{(1)}$, we  can associate the operator
$A^{(k)}~:=~(A^{(1)}\otimes\mathds{1}_d\cdots\otimes\mathds{1}_d)
 + (\mathds{1}_d\otimes A^{(1)}\otimes\cdots\otimes\mathds{1}_d) +\cdots+ (\mathds{1}_d\otimes\cdots\otimes\mathds{1}_d\otimes A^{(1)})$ on $\mathcal{H}^{(k)}$.
The operator $A^{(k)}$ preserves the symmetries of $\mathcal{H}^{(k)}$. The map $A^{(1)} \longrightarrow A^{(k)}$
allows us to study subalgebras of one-particle observables. These constructions are most conveniently expressed in
terms of a \emph{coproduct} $\Delta$ \cite{Balachandran2010}. In fact, an approach based on Hopf algebras
\cite{Balachandran2010} has the advantage that para- and braid-statistics can be \emph{automatically} included. In what
follows we use the simple coproduct $\Delta (g) = g\otimes g$, $g\in U(d)$, linearly extended to all of $\mathds C
U(d)$. It gives the formula for $A^{(k)}$ at the Lie algebra level. Physically, the existence of such a coproduct is
very important. It allows us to homomorphically represent one-particle observables in the $k$-particle sector. In the
examples considered below, observables on such identical-particle systems can also
 be described in terms of creation/annihilation operators.

In the following examples we will concentrate, for the sake of clarity, on systems of two fermions and two bosons (more
examples will be presented in a forthcoming paper). However, our methods can be easily generalized to study
many-particle entanglement.

\textbf{Two Fermions - }
 Consider, as in \cite{Eckert2002},  a one-particle space describing fermions with two \emph{external} degrees of
freedom (e.g. \emph{`left'} and \emph{`right'}) and two \emph{internal} degrees of freedom (e.g. \emph{`spin 1/2'}).
They are described by fermionic creation/annihilation operators $a^{(\dagger)}_\lambda, b^{(\dagger)}_\lambda$, with
$a$ standing for \emph{`left'}, $b$ for \emph{`right'} and $\lambda=1,2$ for spin up and down, respectively. The
single-particle space is therefore $\mathcal H^{(1)}=\mathds C^4$. The two-fermion space is given by $\mathcal
H^{(2)}=\bigwedge^2 \mathds C^3\subset\mathcal H^{(1)} \otimes \mathcal H^{(1)}$ ($\bigwedge$ denoting
anti-symmetrization).  $\mathcal H^{(2)}$ is generated from the ``vacuum'' $|\Omega\rangle$ using pairs of creation
operators. An orthonormal basis is given
 by the  vectors $a^\dagger_1 a^\dagger_2|\Omega\rangle$, $b^\dagger_1
b^\dagger_2|\Omega\rangle$ and $a^\dagger_\lambda b^\dagger_{\lambda'}|\Omega\rangle$, with $\lambda,\lambda'\in\{
1,2\}$. The two-particle algebra $\mathcal A$ of observables is thus isomorphic to the  matrix algebra $M_{6}(\mathds
C)$.

For $|\psi_\theta\rangle=(\cos\theta a_1^\dagger b_2^\dagger + \sin\theta a_2^\dagger b_1^\dagger) |\Omega\rangle$, the
corresponding state $\omega_\theta$ is given by $\omega_{\theta} (\alpha)= \langle \psi_{\theta}
|\alpha|\psi_{\theta}\rangle$ for $\alpha\in \mathcal A$. We now choose the subalgebra $\mathcal A_0$ to be given by
the \emph{one-particle observables corresponding to measurements at the left location}. It is generated by $\mathds
1_{\mathcal A}$, $n_{12}= a_1^\dagger a_1 a_2^\dagger a_2$, $N_a=a_1^\dagger a_1 + a_2^\dagger a_2$ and
$T_{i=1,2,3}=(1/2)~a_\lambda^\dagger (\sigma_i)^{\lambda\lambda'} a_{\lambda'}$. Now we consider the restriction of
$\omega_\theta$ to $\mathcal A_0$ and study the GNS representation corresponding to this choice. For $0<\theta <
\pi/2$, the null space turns out to be spanned by $|n_{12}\rangle $ and $|\mathds 1_{\mathcal A}-N_a\rangle$.
Therefore, the GNS-space $\mathcal H_\theta$ is four-dimensional and spanned by $|[\mathds 1_{\mathcal A}]\rangle$ and
$\{|[T_i] \rangle\}_{i=1,2,3}$. One may show that $\mathcal H_\theta=\mathcal H_1\oplus\mathcal H_2$, with
$\mathcal{H}_1$ spanned by $|[T_1+iT_2]\rangle=|[a_1^\dagger a_2]\rangle$ and $|[a_2^\dagger a_2]\rangle$, and
$\mathcal H_2$ spanned by $|[a_1^\dagger a_1]\rangle$ and $|[T_1-iT_2]\rangle=|[a_2^\dagger a_1]\rangle$. The two
representations are isomorphic. Moreover, from the decomposition $|[\mathds 1_{\mathcal A}]\rangle = |[a_2^\dagger
a_2]\rangle + |[a_1^\dagger a_1]\rangle$ of $|[\mathds 1_{\mathcal A}]\rangle$ into these irreducible subspaces, we
obtain the entropy $S(\theta) = -\cos^2 \theta \log_2 \cos^2\theta - \sin^2 \theta \log_2 \sin^2\theta$.

For $\theta = 0$, the null space is spanned by $|n_{12}\rangle$, $|{\mathds
1_{\mathcal A}} - a_1^\dagger a_1 \rangle$, $|a_2^\dagger a_2\rangle,
|a_1^\dagger a_2\rangle$. The GNS-space $\mathcal H_0$ is $\mathds C^2$ and
isomorphic to the above $\mathcal H_2$. Similarly, for $\theta = \pi/2$ we
find that the GNS-space is isomorphic to the above $\mathcal H_1$. Both
GNS-spaces are irreducible, so that the corresponding $\omega_{0,0}$ and
$\omega_{\frac{\pi}{2},0} $ are pure states with zero entropy.

\emph{This result should  be contrasted with the entropy $S=\log_2 2$ obtained via partial trace  for states with
Slater rank one  such as $\omega_{0,0}$, $\omega_{\frac{\pi}{2},0}$ above (cf. \cite{Ghirardi2004,Tichy2011} and
references therein), that correspond to simple Slater determinants and, therefore, should not be regarded as entangled
states.}

\textbf{Two Bosons - }
Consider the one-particle space $\mathcal H^{(1)}=\mathds C^3$ with an orthonormal basis $\{|e_1\rangle,
|e_2\rangle,|e_3\rangle\}$. The two-boson space $\mathcal H^{(2)}$ is the space of symmetrized vectors in  $\mathcal
H^{(1)}\otimes \mathcal H^{(1)}$. It corresponds to the decomposition $3\otimes 3 = 6\oplus \bar 3$ of $SU(3)$. An
orthonormal basis for $\mathcal H^{(2)}$ is given by vectors $\{|e_i \vee e_j\rangle \}_{i,j\in \{1,2,3\}}$ where
$\vee$ denotes symmetrization (and the vectors are normalized). The two-boson algebra of observables $\mathcal A^{(2)}$
is thus isomorphic to $M_6(\mathds C)$.

For the particular choice $|\psi_{\theta,\phi}\rangle =
\sin\theta\cos\phi|e_1 \vee e_2\rangle +\sin\theta\sin\phi|e_1 \vee
e_3\rangle + \cos\theta|e_3 \vee e_3\rangle$, the corresponding state is
$\omega_{\theta,\phi}$ defined by $\omega_{\theta,\phi} (\alpha)=\langle
\psi_{\theta,\phi} |\alpha| \psi_{\theta,\phi}\rangle$ for any $\alpha\in
\mathcal A$. For the sake
of concreteness, we choose $\mathcal A_0$ to be  given by those
one-particle
observables pertaining \emph{only} to the one-particle states $|e_1\rangle$
and $|e_2\rangle$.

We consider the restriction $\omega_{\theta,\phi}\mid_{\mathcal A_0}$. The
6 representation under the $SU(2)$ action on $|e_1\rangle$ and
$|e_2\rangle$,
splits as $6=3\oplus 2\oplus 1$. The subalgebra $\mathcal A_0$ is given by
block-diagonal matrices. Each block corresponds to one of the irreducible
components in the decomposition $6=3\oplus 2\oplus 1$. The dimension of
$\mathcal A_0$ is therefore $3^2+2^2+1^2 ~=~ 14$.

The construction of the corresponding GNS-representation follows the same procedure as in the previous example. The von
Neumann entropy as a function of  the parameters is given by $S(\theta,\phi) = -\sin^2\theta[\cos^2\phi\log_2 (\sin
\theta \cos\phi)^2 +  \sin^2\phi \log_2 (\sin \theta \sin\phi)^2] - \cos^2\theta\log_2 (\cos\theta)^2\nonumber$.

\section{Conclusions}
The strong point of our approach is that it provides a precise, universal, and mathematically
natural way to characterize and quantify entanglement for systems of
identical particles. For many years it has been known that the von Neumann entropy
based on partial tracing does not give the physically correct answer when
applied to systems of identical particles. Different (i.e. non-universal) criteria
have been developed which strongly depend on the statistics of the particles. In
contrast, our approach is conceptually clear and applies equally to any quantum
system. It thus promises to resolve the controversy regarding entanglement of
identical particles~\cite{Tichy2011}.

\emph{Acknowledgments}: The authors would like to thank Alonso Botero for
discussions that led to this work. ARQ and AFRL  acknowledge the warm
hospitality of  T.R. Govindarajan at The Institute of Mathematical
Sciences, Chennai, where the main part of this work was done. We also thank
M. Asorey, S. Ghosh, K. Gupta, A. Ibort, G. Marmo and V. P. Nair for
fruitful discussions. APB is supported by DOE under grant number
DE-FG02-85ER40231 and by the Institute of Mathematical Sciences, Chennai.
ARQ is supported by CNPq under process number 307760/2009-0. AFRL is
supported by Universidad de los Andes.

\end{document}